\documentclass[sigconf, nofontspec]{acmart}
\usepackage{comment}

\AtBeginDocument{%
  }

\copyrightyear{2026}
\acmYear{2026}
\setcopyright{rightsretained}
\acmConference[HRI'26]{Proceedings of the 2026 ACM/IEEE International Conference on Human-Robot Interaction}{March 16--19, 2026}{Edinburgh, Scotland, UK.} \acmBooktitle{Proceedings of the 2026 ACM/IEEE International Conference on Human-Robot Interaction (HRI'26), March 16--19, 2026, Edinburgh, Scotland, UK} 

\makeatletter
\gdef\@copyrightpermission{
  \begin{minipage}{0.3\columnwidth}
   \href{https://creativecommons.org/licenses/by/4.0/}{\includegraphics[width=0.90\textwidth]{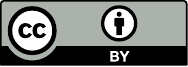}}
  \end{minipage}\hfill
  \begin{minipage}{0.7\columnwidth}
   \href{https://creativecommons.org/licenses/by/4.0/}{This work is licensed under a Creative Commons Attribution International 4.0 License.}
  \end{minipage}
  \vspace{5pt}
}
\makeatother

\begin{document}

\title{Social Robotics for Disabled Students: An Empirical Investigation of Embodiment, Roles and Interaction}

\author{Alva Markelius}
\email{ajkm4@cam.ac.uk}
\orcid{0009-0003-4580-9997}
\affiliation{%
  \institution{University of Cambridge, CST}
 \country{UK}
}

\author{Fethiye Irmak Doğan}
\email{fid21@cam.ac.uk}
\affiliation{
\institution{University of Cambridge, CST}
%
\country{UK}
}

\author{Julie Bailey}
\email{jb658@cam.ac.uk}
\affiliation{
\institution{University of Cambridge, EDUC}
%
\orcid{0000-0002-3773-7546}
 \country{UK}
}

\author{Guy Laban}
\authornote{Contributed due to being a postdoc at Cambridge AFAR Lab during the project.}
\email{laban@bgu.ac.il}
\affiliation{
\institution{Ben Gurion Univ. of the Negev}
\country{Israel}
}

\author{Jenny L. Gibson}
\email{jlg53@cam.ac.uk}
\affiliation{%
\institution{University of Cambridge, EDUC}
 \orcid{0000-0002-6172-6265}
 \country{UK}
}

\author{Hatice Gunes}
\email{hg410@cam.ac.uk}
\affiliation{%
\institution{University of Cambridge, CST}
\orcid{0000-0003-2407-3012}
\country{UK}
}

\renewcommand{\shortauthors}{Markelius et al.}

\begin{abstract}
Institutional and social barriers in higher education often prevent students with disabilities from effectively accessing support, including lengthy procedures, insufficient information, and high social-emotional demands. This study empirically explores how disabled students perceive robot-based support, comparing two interaction roles, one information based (signposting) and one disclosure based (sounding board), and two embodiment types (physical robot/disembodied voice agent). Participants assessed these systems across five dimensions: perceived understanding, social energy demands, information access/clarity, task difficulty, and data privacy concerns. The main findings of the study reveal that the physical robot was perceived as more understanding than the voice-only agent, with embodiment significantly shaping perceptions of sociability, animacy, and privacy. We also analyse differences between disability types. These results provide critical insights into the potential of social robots to mitigate accessibility barriers in higher education, while highlighting ethical, social and technical challenges. Dataset and analysis scripts are available online\footnote{https://github.com/AlvaMarkelius/R4D-dataanalysis}.

\end{abstract}






\maketitle

\section{Introduction}
As of 2022–2023, 15\% of US and 20\% of UK students reported having a disability, a substantial and long-term condition impacting daily life, such as autism, physical or sensory impairment, or mental health condition ~\cite{parliamentSpecialEducational, nces2024students}. Despite existing support mechanisms, many feel unheard or overwhelmed when navigating them. Nearly half of disabled students choose not to raise issues due to lengthy, complex procedures, while another 45\% lack sufficient information about available adjustments~\cite{disabledstudentsuk2023access}. These findings point to a need for clearer, more accessible avenues for addressing concerns, which social robots could help meet by serving as approachable intermediaries that provide timely information, guide students through procedures, and reduce social and institutional barriers to support.

Previous research shed light on the need for disability justice and contextual understanding in using social robots for disability~\cite{dehnert2024ability, haidenhofer2024research}. Although social robots have shown potential for engagement~\cite{Balasuriya_Sitbon_Brereton_Koplick_2020, lalwani2024productivity}, assistance~\cite{Rosenberg-Kima_Koren_Gordon_2020}, communication~\cite{khasawneh2024teacher}, and disclosure~\cite{10700607,LabanGeorgeMorrisonCross+2021+136+159}, little research examines how these interventions align with disabled students' lived experiences. Ensuring that robots are perceived as understanding and responsive to those experiences remains a central challenge (\textbf{Challenge 1}). Another barrier highlighted in prior work is the high social and emotional effort required to engage within institutional contexts \cite{markelius2025stakeholder}. Although social robots have been explored as tools to reduce cognitive and social load by acting as mediators or facilitators~\cite{10.1145/2157689.2157697, kim2013caregiving}, further research is needed to determine whether they can lessen the social energy demands placed on disabled students when navigating tasks in higher education (HE) (\textbf{Challenge 2}). Information access also represents a recurring problem \cite{markelius2025stakeholder}: disabled students frequently report insufficient and unclear guidance on available adjustments~\cite{disabledstudentsuk2023access}, yet little research has assessed whether social robots can address these gaps in clarity and accessibility (\textbf{Challenge 3}). Closely related, navigating layered and abstract institutional processes has been identified as a key accessibility barrier~\cite{disabledstudentsuk2023access}, but few studies consider whether robot-based support could improve task comprehension and completion in these contexts (\textbf{Challenge 4}). Finally, while research consistently identifies privacy as a central ethical concern in educational contexts~\cite{10.1145/3623809.3623816}, and disabled students themselves emphasise it as a non-negotiable condition for adopting robotic support in HE \cite{markelius2025stakeholder}, it remains an open challenge to understand how robot-based interventions shape students' perceptions of privacy in practice (\textbf{Challenge 5}).

Building on these challenges, our work is guided by five research questions. To address concerns of lived-experience alignment, we ask to what extent a robot-based intervention is perceived by students with disabilities as understanding and responsive to their lived experiences (\textbf{RQ1}). In response to the high social effort associated with advocacy and mediation, we examine to what extent a robot-based intervention affects the social energy demands of a mediation task for students with disabilities (\textbf{RQ2}). To mitigate issues of insufficient and unclear information, we investigate to what extent a robot-based intervention can improve information access and clarity for students with disabilities (\textbf{RQ3}). Recognising barriers in navigating layered and abstract institutional processes, we further assess the extent to which a robot-based intervention can facilitate task comprehension and ease of task completion for students with disabilities (\textbf{RQ4}). Given that privacy is a non-negotiable condition for disabled students, we explore to what extent a robot-based intervention can affect students with disabilities’ perceptions of privacy (\textbf{RQ5}). We also examined differences across disability types which allowed us to recognise that disability is complex and heterogeneous and how robot-mediated support may be experienced differently across disability groups.

To address these questions, we build on insights from our previous participatory design study \cite{markelius2025stakeholder}, which framed social robots as a potentially predictable and non-judgemental medium for supporting disabled students' mediation and advocacy in higher education (particularly when navigating information needs and disclosure-related communication), and in which disabled students identified two particularly supportive robot roles: \textit{signposting} for managing information and \textit{sounding board} for managing disclosure. 
We operationalised these in a within-subjects experiment with disabled HE students (N = 31), who interacted with a humanoid robot (Pepper), a voice-only agent (Amazon Echo Dot), which enabled systematic comparison of role and embodiment. While there exists a lot of literature on embodiment, the majority focuses on neurotypical populations. To the authors’ knowledge, there exists no sustained body of work that investigates how the embodied form of a robot is perceived, interpreted, and engaged with by disabled students nor how such embodiment might differently mediate accessibility in interaction. The novelty of this study therefore lies in interrogating embodiment through the lens of disability and accessibility. This is essential, as disability fundamentally alters the perceptual, affective, and social dynamics that underlie HRI \cite{Frennert_Persson_Skavron_2024, Moore_Williams_2021}. Whilst prior work has established that embodiment generally enhances engagement, trust, or enjoyment compared to disembodied agents, such conclusions cannot be uncritically transferred to contexts involving diverse sensory, cognitive, or communicative profiles \cite{Frennert_Persson_Skavron_2024, paterson2024robot}. 

\section{Related Work}
\textbf{Disability Support in Higher Education.} Despite efforts to widen participation in HE, disabled students are still have lower completion rates in higher education than non-disables students \cite{OfS2025StudentCharacteristics}, with studies showing that stigma and concerns around disclosure persist in presenting significant barriers to accessing support, particularly for neurodivergent students \cite{Eccles2018RiskStigma, shaw2024inclusion, Underhill2022AutisticStigma}. Alongside necessary institutional culture change \cite{shaw2024inclusion}, academic success for disabled students is associated with empowerment through self-determination and self-advocacy \cite{Morina2022AcademicSuccess}.  Whilst disabled students point to the need to address basic issues such as physically accessible classrooms \cite{Heffernan2024FailingBasics}, a frustration is the lack of access to prompt lecture materials and recordings \cite{Horlin2024NormalStudent}. Lecture capture plays a key role in supporting the learning of disabled students, both where attendance at lectures is impacted and for neurodivergent students where lecture recordings and transcripts support deeper learning beyond the live lecture \cite{Horlin2024NormalStudent}. However, navigating access to recordings varies considerably, even between departments or courses in the same institution, and can be complex to navigate \cite{Nordmann2022LectureRapture}.

\textbf{Social Robots in Higher Education.} Social robots in HE have been explored in roles including tutors, personalised instructors, assessment facilitators, collaborative partners, and social companions~\cite{Matus_Cano_2025}. For example, Donnermann et al.~\cite{donnermann2022social} implemented an adaptive robotic tutor and reported increased knowledge gain, motivation, and exam performance. A smaller body of work focuses specifically on \textit{students with disabilities}, including robots as schoolwork companions and productivity coaches for students with ADHD~\cite{10.1145/3610977.3634929, lalwani2024productivity}, mental health support via LLM-based CBT and guided breathing~\cite{kian2024can, 9900638}, programming support for students with hearing disabilities~\cite{khasawneh2024teacher}, and non-verbal support for foreign language learning among young adults with learning disabilities~\cite{Vogt_Dunk_Poos_2017}. Despite these efforts, research on social robotics for disabled students in HE remains limited. Notably, user-centred approaches such as participatory and co-design are largely absent, despite being critical for reflecting disabled students’ lived experiences~\cite{haidenhofer2024research}. No studies have examined social robots as mediators of access to information and support, a significant gap given persistent structural barriers including stigma, attainment gaps, and limited social participation~\cite{shaw2024inclusion, brewer2025disabled}.

\textbf{Social Robots for Disability Support.} Social robots have been developed to support people with disabilities across domains including employment~\cite{Takeuchi_Yamazaki_Yoshifuji_2020, Williams_Williams_Moore_McFarlane_2019, Kanetsuna_Takeuchi_Kato_Sono_Osawa_Yoshifuji_Yamazaki_2022}, daily living~\cite{wu_attitudes_2016, dam_experiences_2022}, clinical care~\cite{Shukla_Cristiano_Amela_Anguera_VergésLlahí_Puig_2015, Bertacchini_Demarco_Scuro_Pantano_Bilotta_2023, Sorrentino_Fiorini_Viola_Cavallo_2022}, health coaching~\cite{Robinson_Connolly_Hides_Kavanagh_2020, HunLee_Siewiorek_Smailagic_Bernardino_BermúdezBadia_2023}, non-verbal communication~\cite{Esfandbod_Nourbala_Rokhi_Meghdari_Taheri_Alemi_2024, Fisicaro_Garzotto_Gelsomini_Pozzi_2019, Dennler_Yunis_Realmuto_Sanger_Nikolaidis_Matarić_2021, Oldfield_Broomfield_Philamore_Sewell_Powell_2024}, social companionship~\cite{Bakracheva_Chivarov_Ivanov_2020, Bevilacqua_Maranesi_Felici_Margaritini_Amabili_Barbarossa_Bonfigli_Pelliccioni_Paciaroni_2023, Helm_Carros_Schädler_Wulf_2022}, and social engagement~\cite{DeGroot_Barakova_Lourens_vanWingerden_Sterkenburg_2019, Balasuriya_Sitbon_Brereton_Koplick_2020, markelius2024differential}. These systems typically rely on embodied, affective, or task-based interaction to promote autonomy, accessibility, and well-being~\cite{markscoping}. For instance, voice-controlled robots have supported independent feeding for users with mobility impairments~\cite{Padmanabha_Yuan_Gupta_Karachiwalla_Majidi_Admoni_Erickson_2024}, while conversational robots have enabled sustained engagement in cognitive training for autistic users~\cite{Bertacchini_Demarco_Scuro_Pantano_Bilotta_2023}. Despite this promise, much of the literature adopts a medical model of disability, framing disabled people as in need of `cure'' rather than addressing structural barriers~\cite{Frennert_Persson_Skavron_2024, markscoping}. Critical scholarship highlights how HRI, particularly in autism research, often reproduce dehumanising tropes, including mechanical affect, monotony, and normative assumptions of ability~\cite{Williams_2021, dehnert2024ability}. A large-scale review found widespread pathologisation, essentialised gender and age, and reinforced power asymmetries~\cite{10.1145/3613904.3642798}. Framed as therapeutic or emancipatory, such systems may nevertheless reproduce logics of surveillance, correction, and control rooted in institutional and racialised classification systems~\cite{Nakamura_2019, benjamin2023race, paterson2024robot}. Future research should more critically interrogate the disability models shaping robot design and their implications for agency, power, and inclusion.

\textbf{Social Robots with Multiple Roles.} Prior work shows that social robots can adopt multiple, context-dependent roles. In education, robots have been explored as teachers, assistants, and care receivers~\cite{10.1007/978-3-030-38778-5_38}, with educators viewing them both as didactic tools and social actors~\cite{ekstrom2022dual}. Even when roles are predefined, users may reinterpret them dynamically, highlighting the need for flexible role assignment~\cite{7745213}. Beyond education, robots have similarly taken on diverse roles in wellbeing contexts~\cite{dawe2019can}, including social mediation for children with disabilities~\cite{marti2009robots} and companionship or distraction to reduce loneliness and anxiety~\cite{diaz2016assessing}. Building on these insights and our prior participatory design work~\cite{markelius2025stakeholder}, this study examines two robot roles, \emph{informed guide} and \emph{conversational partner}, mapped to \emph{signposting} and \emph{sounding board} scenarios. To our knowledge, this is the first user study to investigate such roles for social robots supporting disabled students in HE.

\section{The Study}
This study formed the second empirical phase of a broader project on the potential of social robots to support mediation and advocacy for disabled students in HE, building on prior qualitative work with focus groups and interviews \cite{markelius2025stakeholder}. 

\subsection{Study Conditions}
The study employed a 3x2 within-subjects design, with each participant engaging in all six conditions during a single session lasting approximately one hour. The study comprised two roles—signposting and sounding board—delivered through three modes: a physical robot (Pepper), a disembodied voice agent (Amazon Echo Dot), and a control condition. The disembodied agent was included in line with prior work and theory that links embodiment to the transmission of social cues influencing communication fidelity, perception, and self-disclosure \cite{LabanGeorgeMorrisonCross+2021+136+159, li2015benefit}. The laptop control condition served as a status quo control, designed through co-design to reflect participants’ typical experience of completing advocacy- or support-related tasks independently. To mitigate order effects, the sequence of conditions was counterbalanced using a Latin Square design. In each condition, participants were assigned a specific task tailored to the role. For the signposting task, delivered via both the robot (Pepper) and the disembodied voice agent (Amazon Echo Dot), participants were tasked with acquiring information about lecture recordings. In the corresponding control condition, they were asked to find lecture recording information using a laptop and a web browser. For the sounding board task, participants were asked with identifying a personal challenge, discussing it with the agent, and working toward a solution through guided conversation. In the corresponding control condition, participants completed a ``fictive self-referral form,'' designed to mirror a university website form, which prompted the same questions as those posed by the robot. The full self-referral form is in the Supplementary Material.

\subsection{Participants}
\begin{figure}
    \centering
    \includegraphics[width=\linewidth]{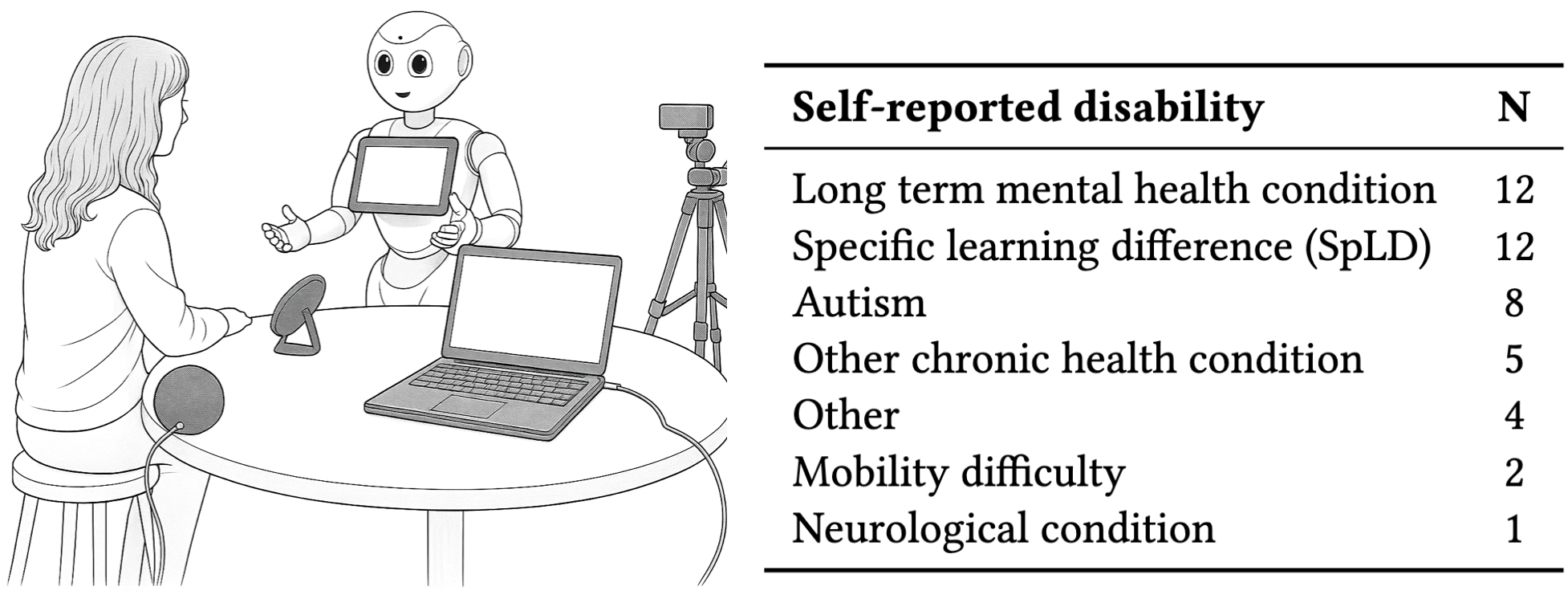}
    \caption{Illustrative figure of the user study setup (left) and number of participants by disability type (right)}
    \label{room}
\vspace{-1.5em}

\end{figure}

A total of 31 disabled students $(N_P = 31)$ from  Cambridge University participated in the user study conducted in March 2025. 
Participants were recruited through student support services, university mailing lists and accessibility networks, and all self-identified as having a disability that impacts their academic experience. The participants were between 19 and 56 years old $(M = 25.2,  SD = 6.7)$. Nine participants identified themselves as male, one as non-binary and the rest as female. Seven of the participants were studying for an undergraduate degree, and the rest were postgraduate students (MS or PhD). Having one or several disabilities was a criterion for inclusion, and the students self-reported their disabilities as seen in Figure \ref{room}, since no universal classification of disabilities exists, we chose 3 nationally standardised options combined to best reflect the characteristics of the participants. The 7 categories listed in Figure \ref{room} represent the disability types reported by participants, there were other options that were not selected. Five of the participants had previous experience of interacting with a social robot, and the rest had not. Informed consent was obtained from all participants, who agreed to the use of their data and the sharing of their anonymised interaction dataset for scientific research purposes. The study design, experimental protocol, and consent materials received approval from the Ethics Committee of Department of Computer Science \& Technology, University of Cambridge. 

\subsection{Experiment Procedure}
The study was conducted in King's College, University of Cambridge. The set-up included a table, chair, microphone (for high-quality audio transcription), and laptop (for completing questionnaires). The robot was placed to the side of the table, while the speaker agent was positioned on the table (see Figure \ref{room}). Each participant attended a single 60-minute session. Upon arrival, participants were welcomed by the experimenter, given time to revisit the information sheet and consent form (previously shared two weeks earlier), and asked to sign the consent form before completing the demographics and pre-study questionnaires. Each participant engaged in three conditions: robot, disembodied agent, and control. Each interaction lasted around five minutes and followed four steps: (1) the agent introduced itself and its role, (2) the student carried out the task with its support, (3) the agent provided assistance according to its role, and (4) the interaction concluded with an invitation for future support. After each condition, participants completed a short five-item questionnaire. Following all six tasks, participants filled out post-study questionnaires and then took part in a semi-structured interview to further explore their experiences (see Section \ref{measures}).

\subsection{Robot Roles}
During the exploratory phase of the project \cite{markelius2025stakeholder}, we identified two complementary roles for social robots in managing information and emotion. In the signposting role, the robot acts as an informed guide, providing tailored, actionable guidance. This encompasses understanding a student’s query or issue, whether academic, social, or administrative, and directing them to appropriate resources or solutions. For instance, the robot might recommend contacting a disability support office, visiting a campus location, accessing online information, or following a specific procedure. The robot might say, ``Would you like me to find your department's policy around lecture recording access?'' or ``Please provide details of the challenge you are currently facing so I can put you in touch with the right person.'' By offering clear, concise, and context-sensitive suggestions, the signposting role ensures students are connected with the accommodations necessary to navigate their challenges. 
The sounding board role instead is a conversational partner that encourages self-reflection and problem-solving by asking guiding questions, offering prompts, or providing a safe space for the student to articulate their thoughts and feelings. It does not directly provide solutions but enables the student’s ability to identify needs and develop strategies. By engaging in active listening and posing open-ended questions, it empowers disabled students to gain clarity, build confidence, and take ownership of their decision-making. For example, the robot might ask, ``What steps have you already considered to address this issue?'' or ``What could be a small part of the issue that you feel comfortable tackling this week?''

\subsection{Robotic and Disembodied Agent Platform}
\textbf{Robotic Platform.} We used the Pepper robot\footnote{https://www.softbankrobotics.com/}, which is a 120 cm tall, semi-humanoid robot. In terms of human-like appearance, Pepper has an anthropomorphic score of 42.2 on the ABOT scale, placing it in the mid-range compared to other humanoid robots\footnote{https://www.abotdatabase.info/}. Pepper was selected based on its prior use in HRI studies on disability~\cite{Mitchell_Sitbon_Balasuriya_Koplick_Beaumont_2021, Fischer-Janzen_Gapp_Götten_Ponomarjova_Blöchle_Wendt_Van, Moore_Williams_2021} and preliminary focus groups with disabled students who identified it as a suitable candidate \cite{markelius2025stakeholder}. Pepper’s physical design is well-suited for semi-public settings such as common areas, student services, and disability resource centres. We used Pepper’s native expressive autonomous voice and gesture repertoire (e.g., nodding while listening and gesturing during greetings) to ensure a consistent interaction. The system employed a hybrid dialogue architecture combining pre-scripted utterances with real-time responses generated via the GPT-4 API. Pre-scripted dialogue followed a predefined flow, while API calls were parameterised with dialogue context (e.g., prior user input and a base system prompt) to enable dynamic responses. This approach balanced flexibility and control, as LLM-generated output can introduce unpredictability that may affect consistency~\cite{Shi_Landrum_O’Connell_Kian_Pinto-Alva_Shrestha_Zhu_Matarić_2024}. To mitigate this, the interaction opening and a set of standardised questions capturing the student’s situation were pre-scripted, ensuring comparable support and reducing the risk of incoherent or inappropriate responses~\cite{irfan2023between, Markelius}. Spoken input was transcribed using the Google Speech Recognition API from audio captured via an external microphone. Pepper’s autonomy was defined following the framework by~\cite{beer2014toward}. In the \textit{sense} stage, the robot operated fully autonomously by capturing and transcribing user speech. In the \textit{plan} stage, pre-scripted dialogue was combined with generative outputs, enabling real-time reasoning while constraining decision-making within a structured dialogue flow. In the \textit{act} stage, Pepper again operated fully autonomously, using built-in text-to-speech and default gestures to enact responses.

\textbf{Agent Platform.} In addition to the robot condition, we implemented the same interaction system on an Alexa Echo Dot device to serve as a voice-only agent platform. The Echo Dot ran the identical underlying codebase used for the robot condition, including the hybrid dialogue system integrating pre-scripted and GPT-4 API-generated responses, as well as speech input transcribed using the Google Speech Recognition API. The only difference between the two platforms was the speech output module: instead of using Pepper’s built-in text-to-speech and gesture system, the Echo Dot condition used the Python pyttsx3 library for voice synthesis. This library supports multiple engines, and in our setup, it was configured to use the default engine available on the host system (e.g., nsss on macOS or sapi5 on Windows). This ensured consistent voice output across participants while maintaining parity in content and interaction logic with the robot condition.

\subsection{Measures}\label{measures}
To address the research questions, we employed a combination of \textit{custom-designed} and \textit{standardised} questionnaires, administered at different stages of the study, together with a brief semi-structured interview. This multi-instrument approach was selected to ensure both direct alignment with the specific context of our research and comparability with validated measures used in prior work. All questionnaires and scales are in the Supplementary Material.

\textbf{Task-level questionnaires.}
The primary instrument was a \textit{custom-designed questionnaire}, developed specifically for this study to ensure that the measures were directly tailored to the research context and the variables of interest identified in earlier participatory work \cite{markelius2025stakeholder}. Existing scales were not used at this stage, as they would not have captured the nuanced concerns and challenges raised by the target group. The questionnaire comprised five items, each mapped to one of the five research questions, and used a five-point Likert scale. It was administered six times, once after each task. The questionnaire items were condition-dependent in two cases. First, the item concerning \textit{information acquisition} was included only in the signposting condition, as the sounding board task did not involve information-seeking and the question would therefore have been inapplicable. Second, the item addressing the \textit{agent's understanding of disability} was excluded from the control condition, as no agent was present and thus participants could not meaningfully evaluate such a dimension. These adjustments were made to ensure the validity and relevance of the questionnaire items across conditions.

\textbf{Pre- and post-questionnaires.}
To complement the task-level measures, participants also completed a short pre- and post questionnaire. These contained five identical items adapted from a \textit{custom-designed questionnaire} used in prior work~\cite{10.1145/3643457, 10.1145/3568162.3577003}. These items assessed participants' general attitudes toward the use of social robots to support students with disabilities, again using a five-point Likert scale. The items covered apprehension, perceived intimidation, perceived usefulness, overall evaluation, and potential impact on mediation, advocacy, and studying. Including both pre- and post-measures allowed us to assess any attitudinal changes over the course of the study.

\textbf{Post-study questionnaires.}
At the end of the study, participants completed two \textit{standardised questionnaires} assessing perceptions of the agents and emotional responses. The Human–Robot Interaction Evaluation Scale (H-REIS)~\cite{spatola2021perception} was administered twice—once for the robot and once for the disembodied agent. It comprises 16 items across four factors (sociability, agency, animacy, and disturbance), rated on a five-point Likert scale, and was included to capture perceptions of anthropomorphism relevant to acceptance of both embodied and disembodied agents. Additionally, the Negative Attitudes toward Robots Scale (NARS)~\cite{nomura2004psychology} was administered for the robot only. NARS includes 14 five-point Likert items measuring anxieties toward everyday interaction with robots and has established reliability and validity, making it suitable for assessing factors influencing adoption in educational contexts.

\textbf{Interviews.}
The quantitative measures were complemented by a brief semi-structured interview at the end of the session. The questions mirrored the items from the pre- and post-questionnaires, but were posed in an open-ended format to encourage participants to elaborate on their responses and reflect more deeply on their experiences of the tasks and of interacting with the robot/agent.

\subsection{Data Analysis}
We adopted a mixed-method approach in which we analysed both quantitative (i.e., standardised and custom-designed questionnaires) and qualitative (semi-structured interviews) data to obtain a comprehensive understanding on social robots for supporting students with disabilities. We analysed the quantitative data using Python statistical libraries\footnote{https://www.statsmodels.org/stable/index.html}. We conducted Shapiro Wilks tests to diagnose normality of the data and decided to conduct non-parametric statistical tests because our samples did not follow a normal distribution. We fit mixed-effects linear models for the task-level questionnaires and for the H-REIS scale, and Wilcoxon signed-rank test for the pre- and post-study questionnaires. We applied iterative reflexive \cite{braun2021thematic} thematic analysis to analyse qualitative data.

\begin{figure}
    \centering
    \includegraphics[width=1\linewidth]{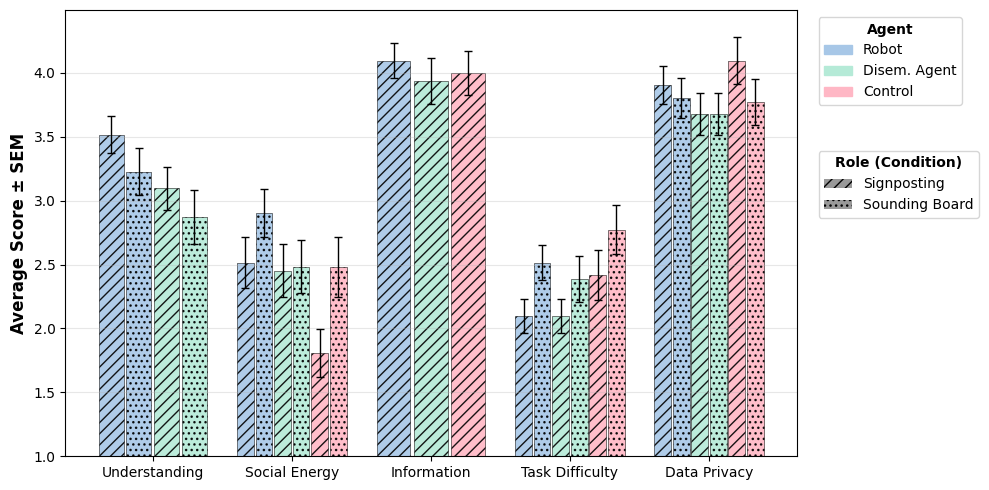}
    \caption{Task-level questionnaire results. Participant ratings of Understanding (RQ1), Social Energy demands (RQ2), Information access and clarity (RQ3), Task Difficulty (RQ4), and Data Privacy concerns (RQ5)}
    \label{fig:betw}
\vspace{-1em}
\end{figure}

\section{Findings}
The first section reports results of the task questionnaires and task-specific questions (between each task) that correspond to each of the research questions as seen in Figure~\ref{fig:betw}. The second section accounts for the pre- and post questions, which were related to attitudes towards social robots for disability. The third section accounts for the thematic analysis of the semi-structured interviews. Supplementary Material contains complete fitted model outputs.

\subsection{Task-level questionnaires}
We fit linear mixed‐effects models separately for the two tasks, \textit{sounding board} and \textit{signposting}, predicting each outcome from \textit{Condition} (robot; disembodied agent; control condition [laptop]), seven disability indicators (Autism, Mobility difficulty, Long‐term mental health condition, Neurological condition, Specific learning difference (SpLD), Other long‐term chronic condition, Other disability), and all Condition~$\times$~Disability interaction terms. As a random effect, we had intercepts for subjects. Significance was calculated using the lmerTest package~\cite{Kuznetsova2017LmerTestModels}, which applies Satterthwaite’s method to estimate degrees of freedom and generate p-values for mixed models. All participants reported at least one disability; disability effects represent differences within this same sample. Random‐effects covariance was singular across models; fixed‐effect inferences remain appropriate, but $R^2_\text{conditional}$ should be interpreted cautiously. 

\textbf{RQ1. Agent Understanding of Disability.}
\textit{Sounding board:} The model explains 77\% of the variance in participants’ perceptions of understanding, whereas the fixed effects in the model explain 35\% of the variance. Despite the variance between participants ($SD = 0.89$), participants reported higher understanding from the \textit{robot} than from the \textit{disembodied agent}, $b=1.16$, $SE=0.38$, $t=3.07$, $p=.002$, 95\%~CI $[0.42,\,1.90]$. This effect was moderated by the disabilities of participants: the effect was smaller for participants with \textit{Autism} ($b=-0.85$, $SE=0.37$, $t=-2.30$, $p=.022$), \textit{Mobility difficulty} ($b=-1.57$, $SE=0.63$, $t=-2.51$, $p=.012$), \textit{Other long‐term chronic condition} ($b=-1.10$, $SE=0.41$, $t=-2.67$, $p=.008$), and participants with \textit{other disability} ($b=-1.49$, $SE=0.58$, $t=-2.59$, $p=.010$). 

\textit{Signposting:} The model explains 58\% of the variance in participants’ perceptions of understanding, whereas the fixed effects in the model explain 32\% of the variance. Participants reported higher understanding from the \textit{robot} than from the \textit{disembodied agent}, $b=0.98$, $SE=0.41$, $t=2.40$, $p=.017$, 95\%~CI $[0.18,\,1.79]$.  This effect was smaller for participants with \textit{Autism} ($b=-0.98$, $SE=0.40$, $t=-2.44$, $p=.015$), \textit{Mobility difficulty} ($b=-1.61$, $SE=0.68$, $t=-2.37$, $p=.018$), and \textit{Other long‐term chronic condition} ($b=-1.32$, $SE=0.44$, $t=-2.96$, $p=.003$).  

\textbf{RQ2. Social Energy}
\textit{Sounding board:} The model explains 32\% of the variance in social-energy demands, whereas the fixed effects explain 25\%. There were no significant differences among conditions. In particular, participants reported similar social-energy demands for the \textit{robot} and the \textit{disembodied agent}, \(b=-0.20\), \(SE=0.69\), \(t=-0.29\), \(p=.769\), 95\%~CI \([-1.55,\,1.15]\). The social energy demands of participating in the \textit{control (laptop)} condition also did not differ from the demands of interacting with the \textit{disembodied agent}, \(b=-0.20\), \(SE=0.69\), \(t=-0.29\), \(p=.774\), 95\%~CI \([-1.55,\,1.15]\), nor from interacting with the \textit{robot}, \(b=-0.00\), \(SE=0.69\), \(t=-0.01\), \(p=.995\), 95\%~CI \([-1.35,\,1.34]\). No Condition~\(\times\)~Disability interactions terms reached significance.

\textit{Signposting:} The model explains 26\% of the variance in social-energy demands, whereas the fixed effects explain 24\%. Between-participant variability was small (\(SD=0.16\)). Relative to interacting with the \textit{disembodied agent}, participating in the \textit{control (laptop)} condition required significantly less social energy, \(b=-1.96\), \(SE=0.69\), \(t=-2.82\), \(p=.005\), 95\%~CI \([-3.32,\,-0.60]\). Interacting with the \textit{robot} also required more social energy than participating in the \textit{control (laptop)} condition, \(b=1.45\), \(SE=0.69\), \(t=2.09\), \(p=.037\), 95\%~CI \([0.09,\,2.81]\). The advantage of the \textit{control (laptop)} condition over the \textit{disembodied agent} was moderated by disability: the contrast was smaller among participants with \textit{Autism} (\(b=1.40\), \(SE=0.68\), \(t=2.06\), \(p=.039\)) and those reporting \textit{Other disability} (\(b=2.29\), \(SE=1.06\), \(t=2.17\), \(p=.030\)). No other Condition~\(\times\)~Disability interaction terms were significant.

\textbf{RQ3. Information Acquisition} 
The model explains 27\% of the variance in perceived information usefulness/clarity, and the fixed effects explain the same proportion; between-participant variance was negligible (\(SD=0.08\); singular random intercept). On average, conditions were similar: participants rated the \textit{robot} and the \textit{disembodied agent} comparably, \(b=0.93\), \(SE=0.55\), \(t=1.67\), \(p=.094\), 95\%~CI \([-0.16,\,2.01]\). Perceptions of the \textit{control (laptop)} did not differ from the \textit{disembodied agent}, \(b=0.38\), \(SE=0.55\), \(t=0.69\), \(p=.492\), 95\%~CI \([-0.70,\,1.47]\); and perceptions of the \textit{robot} did not differ from the \textit{control (laptop)}, \(b=0.55\), \(SE=0.55\), \(t=0.99\), \(p=.324\), 95\%~CI \([-0.54,\,1.63]\). \textit{Autism} moderated the contrast between the \textit{control (laptop)} and the \textit{disembodied agent}: the control was rated lower on information acquisition than the disembodied agent among autistic participants (\( \text{control}\times\text{Autism}: b=-1.33\), \(SE=0.54\), \(t=-2.45\), \(p=.014\), 95\%~CI \([-2.39,\,-0.27]\)). 

\textbf{RQ4. Task difficulty.} \textit{Sounding board:} The model explains 45\% of the variance in perceived task ease/efficiency, whereas the fixed effects explain 36\%. Despite between-participant variability (\(SD=0.29\)), the \textit{disembodied agent} was rated lower than the \textit{control (laptop)} condition, \(b=-1.00\), \(SE=0.50\), \(t=-1.98\), \(p=.048\), 95\%~CI \([-1.99,\,-0.01]\). Participants did not differ in their perceptions of task difficulty when interacting with the \textit{robot} compared to using the \textit{control (laptop)} (\(b=-0.71\), \(SE=0.50\), \(t=-1.41\), \(p=.159\), 95\%~CI \([-1.70,\,0.28]\)). Perceptions of task difficulty also did not differ between interacting with the \textit{robot} and interacting with the \textit{disembodied agent} (\(b=0.29\), \(SE=0.50\), \(t=0.57\), \(p=.568\), 95\%~CI \([-0.70,\,1.28]\)). No Condition\(\times\)Disability interaction terms reached significance. 

\textit{Signposting:} The model explains 26\% of the variance in perceived task ease/efficiency, with the fixed effects accounting for 19\%; between-participant variability was negligible (\(SD=0.14\), singular random intercept). The \textit{disembodied agent} was not significantly different in task difficulty from the \textit{control (laptop)} condition (\(b=-0.28\), \(SE=0.56\), \(t=-0.49\), \(p=.622\), 95\%~CI \([-1.38,\,0.82]\)); and interacting with the \textit{robot} also did not differ from the \textit{control (laptop)} condition (\(b=-0.61\), \(SE=0.56\), \(t=-1.09\), \(p=.276\), 95\%~CI \([-1.71,\,0.49]\)), nor from the \textit{disembodied agent} (\(b=-0.33\), \(SE=0.56\), \(t=-0.60\), \(p=.551\), 95\%~CI \([-1.43,\,0.76]\)). No Condition\(\times\)Disability interactions or baseline disability main effects were significant.

\begin{figure}
    \centering
    \includegraphics[width=\linewidth]{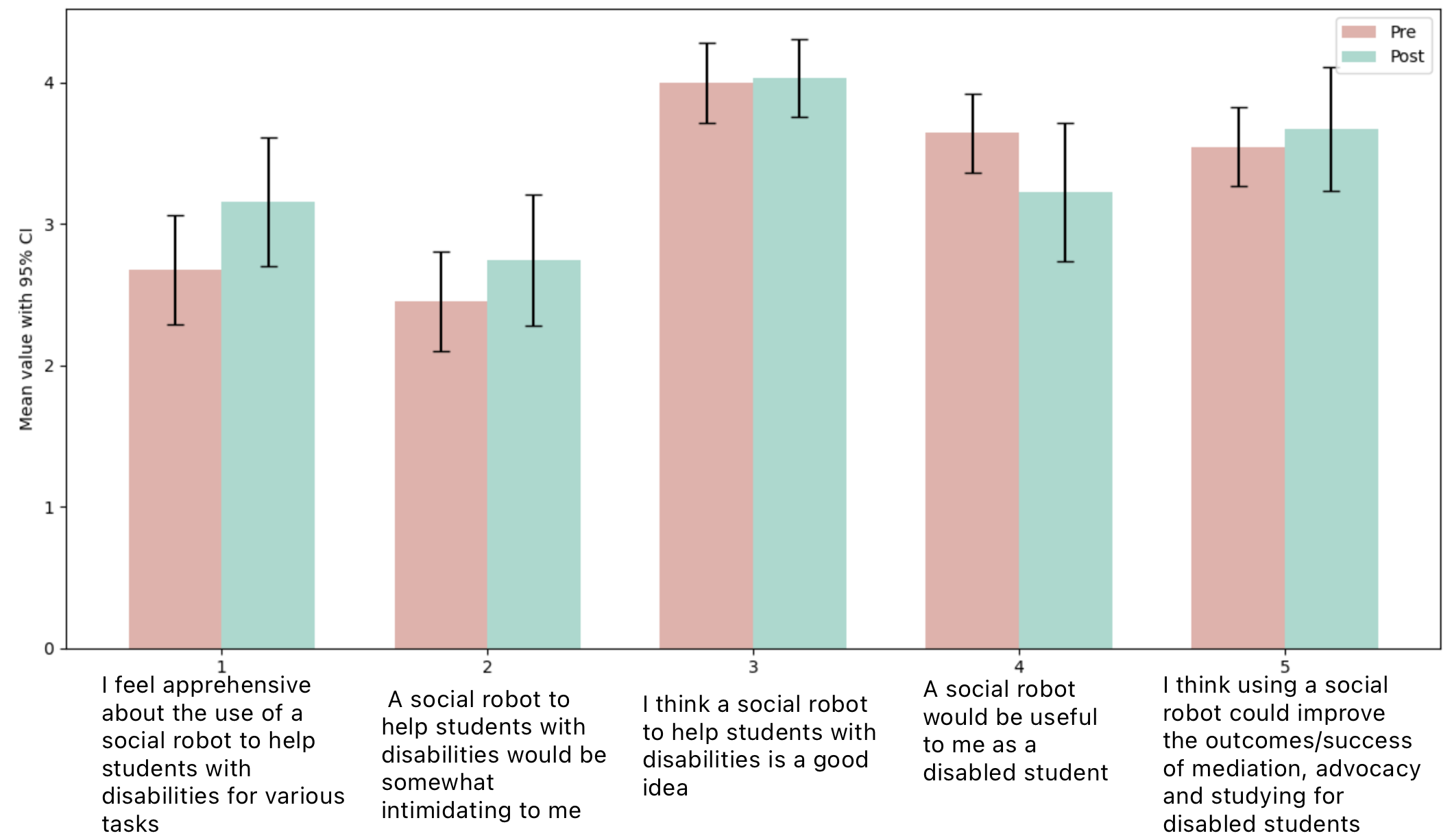}
    \caption{Mean responses to pre/post questionnaire items}
    \label{prepost}
    \vspace{-1em}
\end{figure}

\textbf{RQ5. Data Privacy.} \textit{Sounding board:} The model explains 47\% of the variance in perceived privacy (\(R^2_\text{conditional}=.472\)), whereas the fixed effects explain 28\% (\(R^2_\text{marginal}=.280\)). Despite between-participant variability (\(SD=0.40\)), there were no reliable differences among conditions. Participants rated the \textit{robot} and the \textit{disembodied agent} similarly, \(b=0.80\), \(SE=0.47\), \(t=1.69\), \(p=.091\), 95\%~CI \([-0.13,\,1.72]\); the \textit{control (laptop)} did not differ from the \textit{disembodied agent}, \(b=0.71\), \(SE=0.47\), \(t=1.51\), \(p=.132\), 95\%~CI \([-0.21,\,1.63]\); nor did the \textit{robot} differ from the \textit{control (laptop)}, \(b=0.09\), \(SE=0.47\), \(t=0.18\), \(p=.854\), 95\%~CI \([-0.84,\,1.01]\). No Condition~\(\times\)~Disability interactions reached significance.

\textit{Signposting:} The model explains 53\% of the variance in perceived privacy, whereas the fixed effects explain 28\%; between-participant variability was modest (\(SD=0.46\)). Relative to the \textit{disembodied agent}, the \textit{control (laptop)} condition was perceived as more private, \(b=0.90\), \(SE=0.45\), \(t=1.99\), \(p=.047\), 95\%~CI \([0.01,\,1.78]\). The \textit{robot} did not differ in perceptions of privacy from the \textit{control (laptop)} condition, \(b=-0.41\), \(SE=0.45\), \(t=-0.90\), \(p=.368\), 95\%~CI \([-1.29,\,0.48]\). 

\subsection{Pre/post questionnaires}
The pre- and post questions were related to attitudes towards social robots for disability as follows: i) I feel apprehensive about the use of a social robot to help students with disabilities for various tasks (e.g., sounding board, signposting, study partner). ii) A social robot to help students with disabilities would be somewhat intimidating to me. iii) I think a social robot to help students with disabilities is a good idea. iv) A social robot would be useful to me as a disabled student. v) I think using a social robot could improve the outcomes/success of mediation, advocacy and studying for disabled students. The mean values of the responses can be seen in Figure \ref{prepost}. A Wilcoxon signed-rank test was conducted to compare pre- and post-interaction responses to all questions. For Question 4 (``A social robot would be useful to me as a disabled student''), there was a statistically significant difference between the pre- and post-scores, W = 64.00, p = .034, suggesting that participants rated the usefulness of the robot lower after the interaction. For question 1 (``I feel apprehensive about the use of a social robot to help students with disabilities for various tasks...''), the test revealed a statistically significant increase in apprehension, W = 31.00, p = .012, suggesting that participants felt more apprehensive after the interaction. The other questions did not yield any significance.

\subsection{Post-study questionnaires}
\begin{figure}
    \centering
    \includegraphics[width=0.75\linewidth]{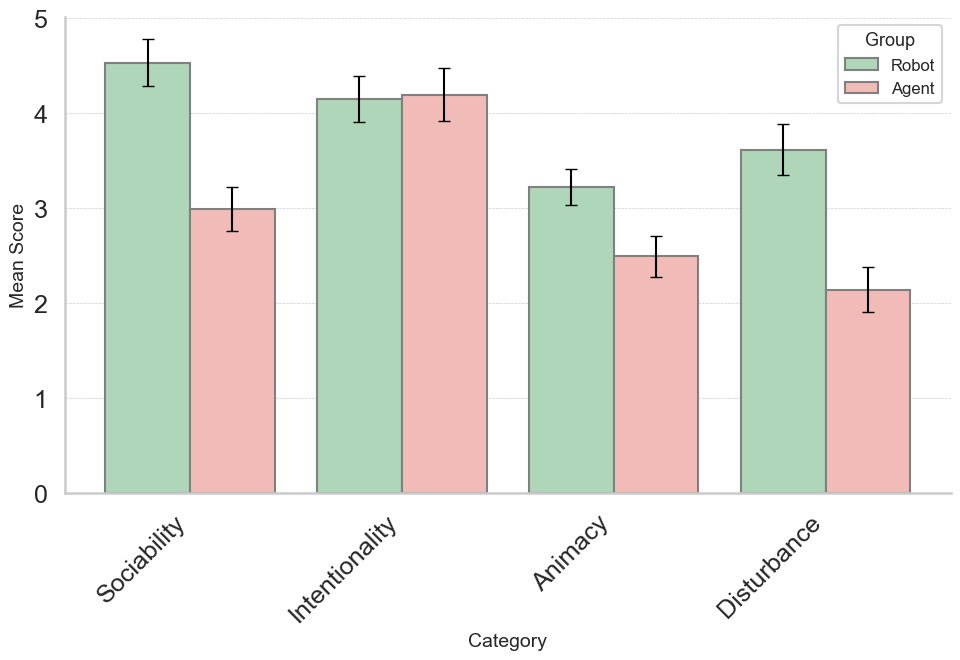}
    \caption{H-REIS scale ratings with ±1 standard error of mean}
    \label{hreis}
    \vspace{-2em}
    \Description{Bar chart showing H-REIS scale ratings with error bars representing plus or minus one standard error of the mean.}
\end{figure}

Additionally, two standardised questionnaires were used to measure attitudes and perceptions of the robot and the disembodied agent. Firstly, the results of the H-REIS scale can be seen in Figure~\ref{hreis}, which was analysed according to its four corresponding categories:
\textbf{Sociability}: Warm, Likeable, Trustworthy, Friendly. \textbf{Intentionality}: Rational, Self-reliant, Intelligent, Intentional. \textbf{Animacy}: Human-like, Real, Natural, Alive. \textbf{Disturbance}: Weird, Creepy, Uncanny, Scary. To compare perceptions of Robot and Agent across four composite categories (Sociability, Intentionality, Animacy, and Disturbance), each category score was computed as the mean of four attribute ratings. We fit linear mixed‐effects models predicting each outcome from \textit{Condition} (robot vs disembodied agent), seven disability indicators (Autism, Mobility difficulty, Long‐term mental health condition, Neurological condition, SpLD, Other long‐term chronic condition, Other disability), and all Condition~$\times$~Disability interaction terms. As a random effect, we had intercepts for subjects. In all models, the \textit{robot} served as the reference; thus, negative condition coefficients indicate lower scores for the \textit{disembodied agent} relative to the \textit{robot}. Significance was calculated using the lmerTest package~\cite{Kuznetsova2017LmerTestModels}, which applies Satterthwaite’s method to estimate degrees of freedom and generate p-values for mixed models. All participants reported at least one disability; disability effects represent differences within this same sample.

\textbf{Sociability.} The model explains 72\% of the variance in participants’ perceptions of sociability, whereas the fixed effects in the model explain 33\% of the variance. Despite the variance between participants ($SD = 1.01$), the disembodied agent was rated less sociable than the robot, $b=-2.25$, $SE=0.433$, $z=-5.20$, $p<.001$, 95\%~CI~$[-3.103,\,-1.404]$. The \textit{condition}~$\times$~\textit{other disability} interaction was significant, $b=2.170$, $SE=0.660$, $z=3.29$, $p=.001$, 95\%~CI~$[0.876,\,3.464]$, indicating that the robot--agent sociability gap was attenuated among respondents reporting other disability.

\textbf{Disturbance.} The model explains 87\% of the variance in participants’ perceptions of disturbance, whereas the fixed effects in the model explain 32\% of the variance. Overall, despite the variance between participants ($SD = 1.27$), the disembodied agent was perceived as less disturbing than the robot, $b=-1.21$, $SE=0.31$, $z=-3.87$, $p<.001$, 95\%~CI~$[-1.829,\,-0.599]$. Although the difference was statistically significant, the robot’s disturbance mean was only 3.6, close to the scale midpoint. This difference was moderated by disability: This effect was stronger for participants with \textit{long-term mental health condition}, $b=-0.859$, $SE=0.287$, $z=-2.99$, $p=.003$, 95\%~CI~$[-1.421,\,-0.296]$, whereas the effect was significantly weaker (i.e., robots still more disturbing than the disembodied agent, but not as much) for participants with \textit{SpLD}, $b=-0.687$, $SE=0.300$, $z=-2.29$, $p=.022$, 95\%~CI~$[-1.276,\,-0.099]$ (subgroup effect $=-1.901$) and for participants with \textit{autism}, $b=0.659$, $SE=0.307$, $z=2.15$, $p=.032$, 95\%~CI~$[0.058,\,1.260]$ (subgroup effect $=-0.555$). Moreover, participants with disability of \textit{mobility difficulty} found the disembodied agent was more disturbing than the robot, $b=2.198$, $SE=0.520$, $z=4.22$, $p<.001$, 95\%~CI~$[1.178,\,3.217]$ (subgroup effect $=0.984$). 

\textbf{Agency.} The model explains 92\% of the variance in participants’ perceptions of agency, whereas the fixed effects in the model explain 11.5\% of the variance. However, there was no significant main effect of condition on agency, $b=-0.041$, $SE=0.243$, $z=-0.17$, $p=.867$, 95\%~CI~$[-0.516,\,0.435]$, and no significant interactions with disability status.\\ 
\textbf{Animacy.} The model explains 78\% of the variance in participants’ perceptions of animacy, whereas the fixed effects in the model explain 29\% of the variance. Despite the variance between participants ($SD = 0.89$), the disembodied agent was rated lower on animacy than the robot, $b=-1.436$, $SE=0.299$, $z=-4.80$, $p<.001$, 95\%~CI~$[-2.023,\,-0.850]$. This difference was moderated by disability: the effect was significantly weaker for participants with \textit{long-term mental health condition} ($b=0.722$, $SE=0.274$, $z=2.64$, $p=.008$, 95\%~CI~$[0.186,\,1.258]$; subgroup effect $=-0.714$) and participants with \textit{other long-term chronic health condition} ($b=1.305$, $SE=0.325$, $z=4.02$, $p<.001$, 95\%~CI~$[0.669,\,1.941]$; subgroup effect $=-0.131$). 

\textbf{NARS-Scale.} In addition to the H-REIS, participants completed the NARS-scale, which assessed general attitudes toward social and emotional aspects of HRI. These results offer a broader picture of participant perceptions, including unease, acceptance, and concerns about emotional attachment or societal implications of robotic integration. On average, participants reported feeling relatively relaxed when talking with robots (M = 3.65, SD = 0.75), yet they also expressed concern that something bad might happen if they relied on robots too much (M = 3.32, SD = 1.25). Unease was comparatively lower when asked about practical use, such as having a job requiring robots (M = 2.06, SD = 1.06) or standing in front of a robot (M = 2.13, SD = 1.09). Attitudes toward emotional aspects were mixed: while some participants felt comforted or open to friendship if robots had emotions (M = 2.87–3.26), others expressed unease at the idea of robots genuinely having emotions (M = 3.29, SD = 1.10). 

\subsection{Thematic Analysis of Interviews}
The post-interaction interviews were analysed using reflexive thematic analysis \cite{braun2021thematic} for deeper insights relevant to the research questions.  Four themes emerged. 

\textbf{Theme 1: It actually listened.}
When reflecting on the differences between the agent and robot for interactions with the same purpose, several students identified a difference in listening quality. Most often this was in terms of feeling listened to, for example, ``you open up more somehow because it feels more present or human'' (P1) and ``It actually listened to what I was trying to say, and the interaction flowed better'' (P2).  Conversely, this was also mentioned in terms of being easier to listen to as ``{[The robot] talks much better than the speaker agent. [The agent is] a bit jarring and hard for me to pay attention to. I find it really monotonous, while this one is a little bit more lifelike, I suppose. It grabs my attention'' (P6).

\textbf{Theme 2: It's less exhausting.}
Students often found it difficult to pinpoint or articulate the specific nature of the difference between interacting with the robot and the agent, instead referring to a general preference for the robot. This pleasantness was directly related to lower social energy demands for some participants, for example, ``[The robot] added emotion to it, just made it less exhausting. Essentially, [the agent] felt more draining, even though they gave us the same information'' (P15).

\textbf{Theme 3: You know it's not human.}
Being close to but not human was an key theme as ``it's not actually real so you can say anything and there will be absolutely no judgment'' (P7) and ``it's just not living and breathing [and] I can just turn it off so it feels a lot more comfortable and safer to work with'' (P10),  facilitating access as you can just ``talk and not worry that you're wasting someone's time'' (P14). The human-like elements were seen to reduce the social effort of the interaction, ``[adding] emotion to it, [...] it felt easier and more natural and like less work '' (P15).

\textbf{Theme 4: It gave me the answer.}
Students trusted the robot responses, as ``they knew what they were talking about [...] I didn't mind the pauses because I knew they were thinking about the answer, and then when they gave me the answer, it was succinct'' (P13).

\section{Discussion}

Our study examined how students with different disabilities perceive robot-based support across two roles (signposting and sounding board) and two embodiments (robot vs. disembodied agent). The findings suggest that embodiment enhances relational outcomes, while experiences of social effort, privacy, and disturbance are influenced by task demands and disability type.

First, the robot was perceived as more understanding of disability than the disembodied agent across both tasks, with qualitative data pointing to stronger feelings of being listened to (RQ1). Indeed, this shows the value of embodiment in enabling social presence and relational support, resonating with prior work on how robots can enhance engagement~\cite{Balasuriya_Sitbon_Brereton_Koplick_2020, lalwani2024productivity} and facilitate communication~\cite{khasawneh2024teacher}. Such effects suggest applications in settings where acknowledgment and validation are central, such as first-contact disability services or reflective conversations, potentially overcoming barriers of stigma disabled students~\cite{shaw2024inclusion}. Notably, the effect was more pronounced for disability types with greater consistency (e.g., SpLD) and weaker for autism and chronic health conditions, which are typically less well understood due to higher variability~\cite{happe2020annual}. However, this relational benefit did not translate into higher scores for information access (RQ3) or task ease (RQ4), despite qualitative reports of increased confidence in the information provided. This suggests that for relatively simple or well-structured tasks, self-guided methods may be equally effective, while for more complex or layered institutional processes (such as where accessing lecture recordings is highly restricted \cite{Horlin2024NormalStudent}), robots may offer more meaningful support in enabling access and comprehension within educational contexts.

Second, we found task-specific differences in social-energy demands across conditions (RQ2). In the sounding-board task, social energy did not differ between the robot, disembodied agent, and control conditions. We reason that the control task of completing a self-referral form may itself provoke a social response: participants anticipate the presence of a future recipient, and thus invest effort in considering how their words will be received and the potential for misunderstandings~\cite{heasman2018perspective}. By contrast, in the signposting task, the control condition required significantly less social energy than either agent. Notably, this was attenuated for participants with autism and for those reporting ``other disabilities''. 
These findings underscore that task type, disability-specific needs, and embodiment all shape how agent-based support is experienced. For instance, while agent involvement can add overhead in routine information-seeking, the disclosure demands of emotion-focused tasks may already dominate the interaction. The thematic analysis further revealed that, within these dynamics, the robot was perceived as more pleasant and less socially draining than the disembodied agent. Building on this sensitivity, one promising direction is to explore streamlined dialogue structures, adaptive interaction length, and alternative modalities (e.g., visual, text-based, or multimodal support) to help provide appropriate assistance without introducing additional social burdens.

Third, privacy concerns varied by task (RQ5). In the sounding-board task, perceptions of privacy were comparable across all conditions, likely because completing a self-referral is inherently personal, foregrounding privacy concerns regardless of medium. By contrast, in the signposting task, the disembodied agent was rated as less private than the control, whereas the robot did not differ. This may reflect participants' familiarity with commercial voice assistants, where privacy risks in information-seeking are more salient and widely discussed. Consistent with prior work highlighting privacy as central to adopting agent-based systems in education~\cite{10.1145/3623809.3623816}, these findings emphasise the importance of evaluating privacy relative to task demands, ensuring transparency, clear privacy controls, and deployment practices that safeguard the needs of diverse students.

Finally, pre- and post-study measures revealed important considerations for how social robots are perceived by students with disabilities. On the one hand, the robot was rated higher on sociability and animacy than the disembodied agent, pointing to the potential of embodiment to foster a stronger sense of social presence. On the other hand, interaction also led to increased apprehension and lower perceived usefulness, suggesting that direct experience sharpened awareness of practical limitations and risks. The robot was also perceived as more disturbing than the disembodied agent, though this varied by disability group: students with mental health conditions reported a weaker aversion to disembodied agents than students with SpLD and/or autism, while students with mobility difficulties found the disembodied agent more disturbing than the robot. Indeed, these differences indicate that disability type plays a critical role in shaping how agent-based systems are received, and the need for careful expectation management and for designing robotic systems that are closely aligned with disabled students' lived experiences. Co-design approaches, in particular, can help ensure that such systems address students' concerns and enable trust in their adoption.

\textbf{Limitations.}
Although some disability categories contained a small number of participants and thus introduced unequal variance, all seven categories were retained in the analysis. This decision was made for comprehensiveness and to avoid disregarding the perspectives of underrepresented groups. While this choice means the results are more exploratory and inductive than strongly generalisable, the inclusion of these data provides rare, valuable insights into a diverse set of disabilities and can inform future research and deployments. Second, the study included custom-designed questionnaires rather than solely standardised instruments. While this ensured sensitivity to the specific concerns raised in earlier participatory work, it reduces direct comparability with existing literature. Third, while the participants were drawn from a single HE institution, the issues raised, such as accessing accommodations and navigating support processes, are widely reported across contexts. Therefore, the findings speak to broader patterns in the experiences of disabled students, even if institutional specifics may vary.

\section{Conclusion \& Future Work}

This study presents the first systematic evaluation of social robots for disability support in HE. By comparing roles and embodiments with 31 disabled students, we identify key opportunities and constraints. Robots can strengthen perceptions of understanding and social presence, but they do not yet outperform self-guided methods for information and task completion, and they raise concerns around effort and privacy. Moving forward, future systems should be more tightly integrated with institutional infrastructures, designed to minimise unnecessary social demands, and equipped with robust privacy protections. Most importantly, development must remain grounded in the lived experiences and needs of students with disabilities, supporting the understanding of individual differences and ensuring that robots reduce existing barriers to access, inclusion, and trust.

\begin{acks}
\noindent\textbf{Funding:} This work was supported by the  University of Cambridge School of Technology Seed Fund. A. Markelius is supported by the Cambridge International Trust Scholarship. F. I. Doğan, G. Laban \& H. Gunes were also supported by the EPSRC/UKRI ARoEQ project (grant ref. EP/R030782/1). G. Laban contributed to this work due to being a postdoc at Cambridge AFAR Lab during the project.
\textbf{Open Access:} For open access purposes, the authors have applied a Creative Commons Attribution (CC BY) licence to any Author Accepted Manuscript version arising. \textbf{Data access:} Raw data related to this publication cannot be openly released due to anonymity and privacy issues. \textbf{Contributions:} Conceptualisation: HG, JG, FID, AM, JB. Methodology: AM, FID, JB, HG. 
Software: AM, FID. Formal analysis: AM, GL, JB. Investigation \& Data Curation: AM. Writing - Original Draft: AM, FID, JB, GL. Writing - Review \& Editing: AM, FID, JB, HG. Visualization: AM. Supervision: HG.  
Project administration: HG, JG, AM. Funding acquisition: HG, JG, FID, AM.

\end{acks}

\bibliographystyle{ACM-Reference-Format}
\bibliography{refs}

\end{document}